\documentclass[aps,prb,twocolumn,groupedaddress,showpacs,amsmath,amssymb]{revtex4}

\usepackage{graphicx}% Include figure files 
\usepackage{dcolumn}% Align table columns on decimal point 
\usepackage{bm}% bold math 

\begin{document}

\title{Large bulk resistivity and surface quantum oscillations in the 
topological insulator Bi$_{2}$Te$_{2}$Se}

\author{Zhi Ren}
\author{A. A. Taskin}
\author{Satoshi Sasaki}
\author{Kouji Segawa}
\author{Yoichi Ando}
\email{y_ando@sanken.osaka-u.ac.jp}

\affiliation{Institute of Scientific and Industrial Research,
Osaka University, Ibaraki, Osaka 567-0047, Japan}

\date{\today}

\begin{abstract}

Topological insulators are predicted to present novel surface transport
phenomena, but their experimental studies have been hindered by a
metallic bulk conduction that overwhelms the surface transport. We show
that a new topological insulator, Bi$_{2}$Te$_{2}$Se, presents a high
resistivity exceeding 1 $\Omega$cm and a variable-range hopping
behavior, and yet presents Shubnikov-de Haas oscillations coming from
the surface Dirac fermions. Furthermore, we have been able to clarify
both the bulk and surface transport channels, establishing a
comprehensive understanding of the transport in this material. Our
results demonstrate that Bi$_{2}$Te$_{2}$Se is the best material to date
for studying the surface quantum transport in a topological insulator.

\end{abstract}


% 73.25.+i 	Surface conductivity and carrier phenomena
% 71.18.+y 	Fermi surface: calculations and measurements; effective mass,
%                     g factor
% 73.20.At 	Electron states at surfaces and interfaces - Surface states, 
%                 band structure, electron density of states
% 71.70.Di 	Landau levels
% 71.55.Ht      Impurity and defect levels - Other nonmetals
% 72.20.My      Conductivity phenomena in semiconductors and insulators - 
%                 Galvanomagnetic and other magnetotransport effects 
% 72.20.Ee      Mobility edges; hopping transport  

\maketitle

The three-dimensional (3D) topological insulator (TI) is characterized
by a non-trivial $Z_2$ topology \cite{K2,MB} of the bulk wave function,
and it represents a new topological quantum state realized in a band
insulator. In theory, 3D TIs are insulating in the bulk and unusual
metallic surface states consisting of spin-filtered Dirac fermions give
rise to interesting surface transport phenomena. \cite{Kane,Moore,Zhang}
In reality, however, TI samples available today are invariably
conducting in the bulk, and charge transport is always dominated by the
bulk current.
\cite{Checkelsky,Analytis,Butch,Eto,Ong2010,Taskin1,Taskin2,Peng}
Therefore, to exploit the novel surface transport properties of
topological insulators, it is crucial to achieve a bulk-insulating state
in a TI material. 

Among the recently discovered TIs, Bi$_{2}$Se$_{3}$ has been the most
attractive because of its simple surface-state structure. \cite{Kane}
Unfortunately, near-stoichiometric Bi$_{2}$Se$_{3}$ is always $n$-type
owing to a large amount of Se vacancies. An isostructural material
Bi$_{2}$Te$_{3}$ can be grown as $p$-type, \cite{Ong2010} but usually it
is also highly metallic, most likely due to anti-site defects which are
promoted by close electronegativities of Bi and Te. Significant efforts
have been made \cite{Checkelsky,Analytis,Butch,Eto,Ong2010} to achieve
bulk insulating behavior in Bi$_{2}$Se$_{3}$ and Bi$_{2}$Te$_{3}$;
however, while an increase in resistivity with decreasing temperature
has been observed, so far the bulk remains to be essentially a metal and
a clearly insulating temperature dependence, such as the variable-range
hopping (VRH) behavior, \cite{Shklovskii} has never been reported. For
example, by growing a Bi$_{2}$Te$_{3}$ single crystal with a
compositional gradient, it was possible to observe a resistivity upturn
at low temperature and to measure the surface quantum transport,
\cite{Ong2010} but the resistivity remained low ($<$ 12 m$\Omega$cm) in
absolute terms and the surface contribution to the transport did not
exceed $\sim$0.3\%. \cite{Ong2010} The situation is essentially the same
\cite{Checkelsky,Analytis,Butch,Eto} in Bi$_{2}$Se$_{3}$.

\begin{figure}\includegraphics*[width=8.7cm]{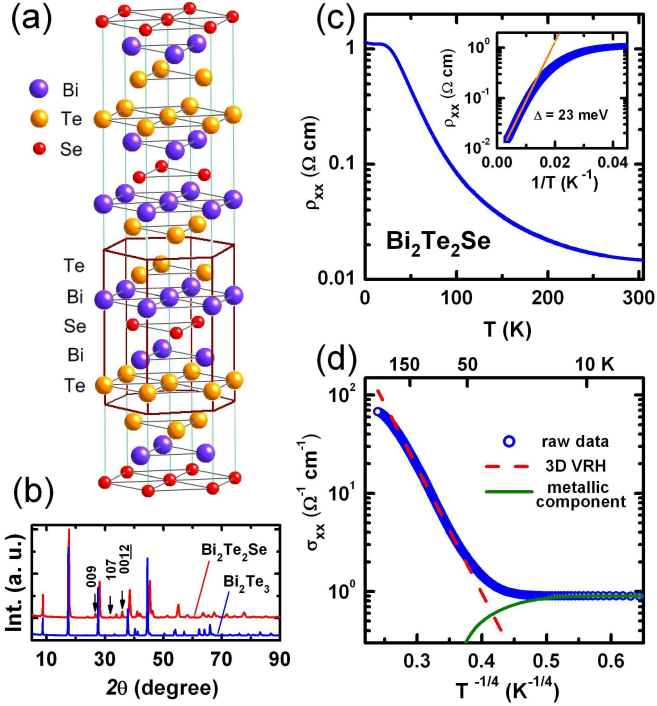}
\caption{(Color online) 
(a) Layered crystal structure of Bi$_{2}$Te$_{2}$Se showing the
ordering of Te and Se atoms.
(b) Comparison of the X-ray powder diffraction patterns of
Bi$_{2}$Te$_{2}$Se and Bi$_{2}$Te$_{3}$. Arrows indicate the peaks
characteristic of Bi$_{2}$Te$_{2}$Se.
(c) Temperature dependence of $\rho_{xx}$; inset shows the
Arrhenius plot.
(d) Plot of the conductivity $\sigma_{xx}$ ($= \rho_{xx}^{-1}$) vs.
$T^{-1/4}$. Dashed line is the fitting of the 3D VRH behavior
$\sigma_{xx} \sim \exp[-(T/T_0)^{-1/4}]$ to the data; deviation
from the fitting at low temperature, shown by solid line, 
signifies the parallel metallic conduction. 
} 
\label{fig1}
\end{figure}

Given this difficulty, searching for a new TI material better suited for
achieving the bulk insulating state is obviously important. In this
paper, we report that a new TI material, Bi$_{2}$Te$_{2}$Se, which has
an ordered tetradymite structure \cite{BTS221} [Fig. 1(a)] with the
basic quintuple-layer unit of Te-Bi-Se-Bi-Te and was recently confirmed
to have a topological surface state, \cite{Xu} has desirable
characteristics for surface transport studies. We found that
high-quality single crystals of ordered Bi$_{2}$Te$_{2}$Se show a high
resistivity exceeding 1 $\Omega$cm, together with a variable-range
hopping (VRH) behavior which is a hallmark of an insulator; yet, it
presents Shubnikov-de Haas (SdH) oscillations which signify the 2D
surface state consistent with the topological one observed by the
angle-resolved photoemission spectroscopy (ARPES). \cite{Xu}

By examining the difference in the doping chemistry between
Bi$_{2}$Se$_{3}$ and Bi$_{2}$Te$_{3}$, one may understand that the
ordered Bi$_{2}$Te$_{2}$Se has reasons to be superior: i) The formation
of Se vacancies is expected to be suppressed, because the Se trapped
between two Bi atoms is less exposed to evaporation due to stronger
chemical bonding with Bi in this position; ii) The formation of the
anti-site defects between Te and Bi is also expected to be suppressed
because of preferable bonding between Se and Bi in contrast to Se-Te
bonding; iii) Ordered nature minimizes the additional disorder that
could be caused by Se/Te randomness. In this work, single crystals of
Bi$_{2}$Te$_{2}$Se were grown by melting high purity (6N) elements of
Bi, Te and Se with a molar ratio of 2:1.95:1.05 at 850$^{\circ}$C for
two days in evacuated quartz tubes, followed by cooling to room
temperature over three weeks. The ordering of the chalcogen layers in
our Bi$_{2}$Te$_{2}$Se single crystals is confirmed by the X-ray powder
diffraction patterns by comparing those from Bi$_{2}$Te$_{2}$Se and
Bi$_{2}$Te$_{3}$ as shown in Fig. 1(b), where the characteristic peaks,
which are intensified in the ordered Bi$_{2}$Te$_{2}$Se compound,
\cite{BTS221} are indicated by arrows. Note that the ordering does not
cause a doubling nor a symmetry change of the unit cell.

For transport measurements, cleaved crystals were aligned using the
X-ray Laue analysis and cut along the principal axes, and the (111)
surface was protected by depositing an alumina thin film after cleaning
the surface by bias sputtering with Ar ions. Ohmic contacts were
prepared by using room-temperature cured silver paste. The sample
reported here was 0.51-mm wide and 0.26-mm thick, with the voltage
contact distance of 0.86 mm. The resistivity $\rho_{xx}$ and the
Hall resistivity $\rho_{yx}$ were measured simultaneously by a standard
six-probe method by sweeping the magnetic field between $\pm$14 T,
during which the temperature was stabilized to within $\pm$5 mK. The
sweep rate was 0.3 T/min.

Figure 1(c) shows the temperature dependence of the resistivity
$\rho_{xx}$ of this ordered compound. We observe almost two orders of
magnitude increase in $\rho_{xx}$ upon cooling from room temperature,
which is an indication of an insulating behavior. Indeed, the Arrhenius
plot [inset of Fig. 1(c)] shows an activated temperature dependence in
the range from 300 K down to $\sim$150 K with an excitation energy
$\Delta$ of about 23 meV. Below $\sim$150 K, the transport is understood
as a parallel circuit of an insulating component characterized by a
3D VRH behavior \cite{Shklovskii} and a metallic component that
saturates below 10 K [Fig. 1(d)].

The saturation of the resistivity at low temperature implies a finite
metallic conductivity at $T$ = 0 K. To clarify the nature of this
metallic state, we employed the SdH oscillations, whose angular
dependence in tilted magnetic fields can provide the information about
the size and the shape of the Fermi surface (FS) and, more importantly,
about the dimensionality of the FS. In our Bi$_{2}$Te$_{2}$Se crystals,
we observed SdH oscillations in both $\rho_{xx}$ and $\rho_{yx}$ 
(the latter presenting more pronounced oscillations), and the two show
essentially the same frequency with a phase shift of approximately
$\pi$.
\cite{phase} Figure 2(a) shows the derivative of $\rho_{yx}$ with
respect to the magnetic field $B$ plotted in 1/$B_{\bot}$ = 1/($B
\cos\theta$) coordinates, where $\theta$ is the angle between $B$ and
the $C_{3}$ axis as shown schematically in the bottom inset. Two
important features can be readily recognized: First, d$\rho_{yx}$/d$B$
is periodic in the inverse magnetic field, indicating the existence of a
well-defined FS. Second, oscillations depend solely on $B_{\bot}$,
implying a 2D character. Note that the oscillations quickly disappear
with increasing $\theta$ (above $\sim$40$^{\circ}$ they are hardly
distinguishable) because the amplitude of the oscillations strongly
depends on the magnetic-field strength. The signature of a 2D FS can be
also seen in the plot of the oscillation frequency vs. $\theta$, which
shows the characteristic $1/\cos \theta$ dependence [upper inset of Fig.
2(a)].

The obtained frequency for $\theta \simeq$ 0$^{\circ}$ of 64 T, which
is directly related to the Fermi-surface cross section $A$ via the
Onsager relation $F = (\hbar c / 2\pi e)A$, gives $k_{F}$ =
4.4$\times$10$^{6}$ cm$^{-1}$, which means the surface charge-carrier
concentration $N_s$ = $k_{F}^{2}/4 \pi$ = 1.5$\times$10$^{12}$ cm$^{-2}$
for a spin-filtered surface state. It is important to notice that our
measured $k_{F}$ is too large if the oscillations come from the bulk;
for example, the carrier concentration for a 3D ellipsoidal FS that
might be consistent with our SdH data would be $\ge$ 1$\times$10$^{19}$
cm$^{-3}$, which is orders of magnitude larger than what we obtain from
the Hall data described later.

\begin{figure}\includegraphics*[width=8.7cm]{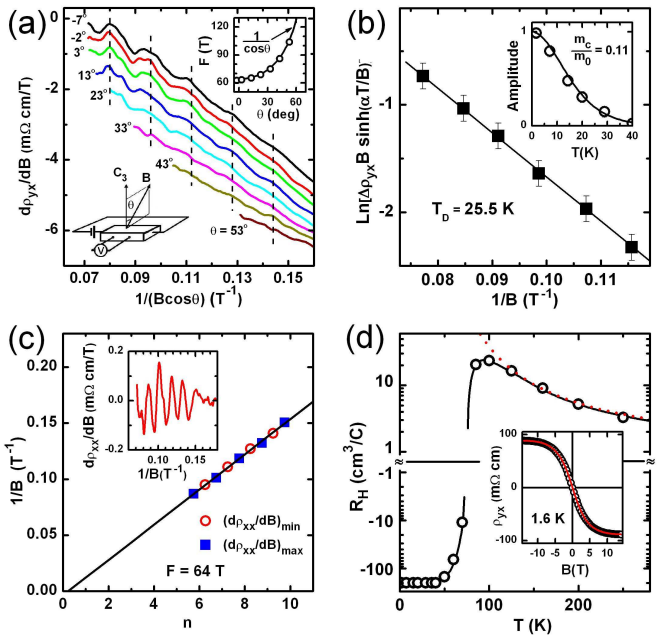}
\caption{(Color online) 
(a) d$\rho_{yx}$/d$B$ vs. 1/$B_{\bot}$ [= $1/(B\cos\theta)$]
in magnetic fields tilted from the $C_{3}$ axis at different
angles $\theta$, where all curves are shifted for clarity; lower inset
shows the schematics of the experiment and the definition of $\theta$.
Equidistant dashed lines in the main panel emphasize that the 
positions of maxima for any $\theta$ depends only on $B_{\bot}$;
upper inset shows the $1/\cos\theta$ dependence of the oscillation
frequency. Both point to the 2D FS.
(b) Dingle plot for the oscillations in $\Delta\rho_{yx}$, which is 
obtained after subtracting a smooth background from $\rho_{yx}$,
giving $T_{D}$ = 25.5 K; inset shows the $T$ dependence of the
SdH amplitude for $\theta \simeq$ 0$^{\circ}$, yielding $m_c$ = 
0.11$m_{e}$.
(c) Landau-level fan diagram for oscillations in $d\rho_{xx}/dB$
measured at $T$ = 1.6 K and $\theta \simeq$ 0$^{\circ}$. Inset shows 
$d\rho_{xx}/dB$ vs. 1/$B$ after subtracting a smooth
background. Minima and maxima in $d\rho_{xx}/dB$ 
correspond to $n+\frac{1}{4}$ and $n+\frac{3}{4}$, respectively. 
The linear fit intersects the axis at $n$ = 0.22.
(d) Temperature dependence of the low-field $R_H$; dotted line 
represents the activated behavior. Inset shows the $\rho_{yx}(B)$ curve 
at 1.6 K and its fittings with the two-band model.
}
\label{fig2}
\end{figure}

Fitting the standard Lifshitz-Kosevich theory \cite{Shoenberg1984} to
the temperature dependence of the SdH amplitudes [inset of Fig. 2(b)]
gives the cyclotron mass $m_{c}$ = 0.11$m_{e}$, where $m_{e}$ is the
free electron mass. Assuming that the electrons are Dirac-like, one
obtains the Fermi velocity $v_{F}$=$\hbar k_{F}$/$m_{c}$ =
4.6$\times$10$^{7}$ cm/s. This $v_{F}$ is consistent with the ARPES
data, \cite{Xu} affirming the Dirac-fermion assumption. The position of
the surface Fermi level $E_{F}^{s}$ measured from the Dirac point can be
estimated from $v_{F}$ and $k_{F}$ to be 130 meV, suggesting that the
observed surface carriers are electrons.

Once $m_c$ is known, the Dingle analysis [shown in Fig. 2(b)] uncovers
the scattering time $\tau$ through the Dingle temperature $T_D$ [=
$\hbar / (2 \pi \tau k_{B}$)]; the slope of the linear fit to the data
yields $T_D$ of 25.5 K, which gives $\tau$ = 4.8$\times$10$^{-14}$ s.
Hence, one obtains the mean free path on the surface $\ell_{s}^{\rm
SdH}$ = $v_{F} \tau \approx$ 22 nm and the surface mobility
$\mu_{s}^{\rm SdH}$ = ($e \ell_{s}^{\rm SdH}$)/($\hbar k_{F}$) $\approx$
760 cm/Vs. Note that both quantities are underestimated, because $\tau$
obtained from the SdH effect reflects scattering events in all
directions, whereas in the transport properties the backward scattering,
which is prohibited in topological insulators, \cite{Kane} plays the
most important role.

In the SdH oscillations, the resistivity oscillates as $\Delta\rho_{xx}
\sim \cos[2\pi(\frac{F}{B}+\frac{1}{2}+\beta)]$, where $2\pi\beta$ is
the Berry phase. \cite{Kim2005} It is known \cite{Kim2005, Mikitik1999,
Niu2010, TAndo} that the cyclotron orbit of an electron in a magnetic
field acquires a Berry phase $\pi$ if its energy dispersion is linear
near the degenerate point (called Dirac point), whereas in ordinary
metals $\beta$ = 0. Experimentally, the phase of the oscillations can be
obtained from the Landau-level fan diagram, which is shown in Fig. 2(c)
for the oscillations in $d\rho_{xx}/dB$. \cite{FanDiagram} The values of
1/$B$ corresponding to minima and maxima in $d\rho_{xx}/dB$ [shown in
Fig. 2(c) inset] are plotted as a function of the Landau level number
$n$. \cite{LLIndex} The extrapolation of the linear fit to the data
shown in Fig. 2(c) suggests $\beta$ = 0.22; however, since our data are
far from the origin, this extrapolation involves a large uncertainty and
we can only conclude that our data suggest a finite Berry phase and are
not inconsistent with the Dirac nature of the surface state. We note
that a clear $\pi$ Berry phase has never been observed in TIs.
\cite{Analytis,Ong2010,Taskin2}

To elucidate the bulk contribution to the transport properties, the Hall
data is useful. As shown in Fig. 2(d), the low-field Hall coefficient
$R_{H}$ (= $\rho_{yx}$/$B$ near $B$ = 0 T) changes sign from positive to
negative upon cooling, signifying the change of dominant charge carries
from holes to electrons. At high temperature ($\gtrsim$ 150 K), the
behavior of $R_{H}(T)$ is thermally activated; the dotted line is an
Arrhenius-law fitting, which signifies the activation of holes with an
effective activation energy $\Delta^{*}$ = 33 meV. This $\Delta^{*}$ is
of the same order as we found for the resistivity [inset of Fig. 1(c)],
and the small difference is attributed to the temperature dependence of
the mobility. The prefactor of the activated behavior gives an estimate
of the acceptor concentration $N_a \simeq$ 9$\times$10$^{18}$ cm$^{-3}$.
\cite{note}

When we look at the magnetic-field dependence of $\rho_{yx}$ at 1.6 K,
the low-field and high-field slopes are essentially different [inset of
Fig. 2(d)]. The low-field $R_{H}$ is $-$200 cm$^{3}$/C, giving the
effective carrier concentration of 3.1$\times$10$^{16}$ cm$^{-3}$. As we
know from SdH oscillations, the surface-electron concentration is
1.5$\times$10$^{12}$ cm$^{-2}$, corresponding to an effective 3D Hall
coefficient of $-5.4\times10^{4}$ cm$^{3}$/C, which is much larger than
the observed value. Therefore, surface electrons alone cannot account
for the low-temperature $R_{H}$ and there should be other charge
carriers in the system. Their concentration and type can be inferred
from the high-field slope of $\rho_{yx}(B)$, since the high-field limit
of $R_H$ is determined only by the number (and type) of carriers
irrespective of their mobilities; in our data, the high-field slope is
$-$26 cm$^{3}$/C, which points to the existence of bulk electrons with
the concentration $n$ of 2.4$\times$10$^{17}$ cm$^{-3}$ in addition to
the surface electrons. This example demonstrates that estimations of the
bulk carrier density based on the low-field $R_{H}$, though often done
in TIs, \cite{Checkelsky,Analytis,Butch} can be too optimistic.

The above analysis indicates that at low temperature the surface
electrons and bulk electrons are contributing in parallel. It turns out
that the standard two-band model \cite{Ashcroft} 
\begin{equation}
\rho_{yx} = \frac{(R_{s}\rho_{n}^{2}+R_{n}\rho_{s}^{2})B+R_{s}R_{n}(R_{s}+R_{n})B^{3}}
{(\rho_{s}+\rho_{n})^{2}+(R_{s}+R_{n})^{2}B^{2}},
\end{equation}
fits the whole $\rho_{yx}(B)$ curve very well [solid line in the inset
of Fig. 2(d)]; here, $R_{n}$ and $\rho_{n}$ are the Hall coefficient and
resistivity of the bulk electrons, $R_{s} = t/(eN_{s})$ and
$\rho_{s}=\rho_{\square}t$, with $\rho_{\square}$ the surface sheet
resistance and $t$ the sample thickness. This fitting yields the
surface mobility $\mu_{s}$ = 1450 cm$^{2}$/Vs and the bulk mobility
$\mu_{n}$ = 11 cm$^{2}$/Vs, along with $N_s$ = 1.5$\times$10$^{12}$
cm$^{-2}$ and $n$ = 2.4$\times$10$^{17}$ cm$^{-3}$. The obtained $\mu_s$
is about two times larger than the $\mu_{s}^{\rm SdH}$ estimated from
the SdH analysis [Fig. 2(b)]. This is expected for the topological
surface state as we mentioned before. On the other hand, the bulk
mobility $\mu_{n}$ of 11 cm$^{2}$/Vs is near the boundary between the
band and hopping transport regimes, \cite{Shklovskii} suggesting that
the bulk electrons move in a very disordered potential. The fraction of
the surface contribution in the total conductance at 1.6 K is calculated
to be $\sim$6\%, which is about 20 times larger than that achieved
\cite{Ong2010} in Bi$_2$Te$_3$.

\begin{figure}\includegraphics*[width=4cm]{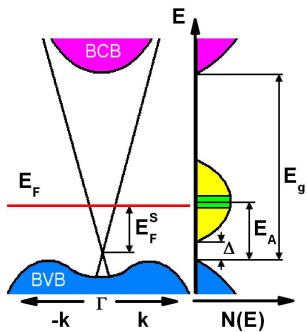}
\caption{(Color online) 
Schematic picture of the bulk and surface band structures (left), 
together with the energy diagram of the density of states of the bulk 
and impurity bands (right). In the impurity band which is due to the 
acceptor levels and is located within the 
band gap, only the central part forms the extended states and the tails
consist of localized states.
} 
\label{fig3}
\end{figure}

From the detailed information we have gathered, a comprehensive picture
emerges for the bulk transport mechanism in Bi$_{2}$Te$_{2}$Se: Taking
into account the relatively large concentration (9$\times$10$^{18}$
cm$^{-3}$) of acceptors, it is reasonable to assume that those acceptors
form an impurity band (IB) within the energy gap of Bi$_{2}$Te$_{2}$Se;
note that the IB is formed because the wave functions of the electrons
bound to impurity sites overlap to form extended states, which are
responsible for the finite $n$-type bulk conduction at zero temperature
with $n$ = 2.4$\times$10$^{17}$ cm$^{-3}$.  The chemical potential is
pinned to this IB at low temperature and is obviously located within the
bulk energy gap (Fig. 3).
As temperature increases, hopping conduction of electrons using
localized states becomes possible, giving rise to the VRH behavior which
occurs in parallel with the degenerate IB conduction. At even higher
temperature, thermal activation of electrons from the bulk valence band
to the acceptor levels takes place, leading to the $p$-type bulk
conduction.

In conclusion, we have shown that Bi$_{2}$Te$_{2}$Se is the best
material to date for studying the surface quantum transport in a TI. The
surface contribution in the total conductance of our Bi$_{2}$Te$_{2}$Se
crystal is $\sim$6\%, which is the largest ever achieved in a TI. The
bulk mobility of 11 cm$^{2}$/Vs indicates that the IB conduction in our
sample is at the verge of localization, and not much further reduction
in the number of acceptors would be needed to quench the degenerate bulk
conduction. Once such crystals become available, they will allow us to
study the plethora of topological quantum phenomena that have been
predicted \cite{Kane,Moore,Zhang} for this class of materials.

\begin{acknowledgments} 

We thank L. Fu, M.Z. Hasan, N.P. Ong, and D. Vanderbilt for helpful
discussions. This work was supported by JSPS (KAKENHI 19674002), MEXT
(KAKENHI 22103004), and AFOSR (AOARD 10-4103).

\end{acknowledgments}


\begin{thebibliography}{99}
\bibitem{phase}
The observed phase shift of $\sim\pi$ is at odds with the expected
$\frac{\pi}{2}$ shift. \cite{Ong2010} The reason for this unexpected
observation is not clear at the moment.
\bibitem{FanDiagram}
Since the oscillations in $\rho_{yx}$ present the unusual $\pi$ phase
shift compared to those in $\rho_{xx}$, the phase analysis of $\rho_{yx}$
is not very meaningful and we concentrate on $\rho_{xx}$ here.
\bibitem{LLIndex}
Since $d\Delta\rho_{xx}/dB \sim 2\pi \frac{F}{B^2}
\sin[2\pi(\frac{F}{B}+\frac{1}{2}+\beta)]$, we assign $n+\frac{1}{4}$
to the minima (and $n+\frac{3}{4}$ to the maxima) in $d\rho_{xx}/dB$.
\bibitem{note}
In the extrinsic range, $R_{H}(T)$ is determined by $\Delta^{*}$ and
$N_a$ as $R_{H}^{-1} \approx e
\sqrt{N_{a}N_{V}}\exp(-\Delta^{*}/k_{B}T)$, where $N_{V}=2(2\pi
m_{p}^{*}k_{B}T/h^{2})^{3/2}$ is the effective density of states of the
valence band with $m_{p}^{*}$ the effective density-of-state mass.
$m_{p}^{*}$ is not known for Bi$_{2}$Te$_{2}$Se, but if we take
$m_{p}^{*} = 0.106 m_e$ measured for Bi$_{2}$Te$_{3}$,
\cite{Kohler1976} $N_{V}$ = 5.2$\times$10$^{18}$ cm$^{-3}$ at 300 K and
$N_{a}$ is estimated to be about 9$\times$10$^{18}$ cm$^{-3}$.
\end{thebibliography}
\end{document}